\documentclass[aps,prd,floatfix,preprintnumbers,letterpaper,ifmbe,twocolumn,usegraphicx]{revtex4-1}
\usepackage{amsfonts,amsmath,amsfonts,amscd}

\begin{document}

\title{On Beltrami states near black hole event horizon}

\author{Chinmoy Bhattacharjee$^{\,1}$\thanks{Corresponding author.}}
\email{usa.chinmoy@gmail.com}

\author{Justin C. Feng$^{\,2}$}

\affiliation{$^{1}$ Department of Physics, New York Institute of Technology, Old Westbury, NY 11586 USA }
\affiliation{$^{2}$CENTRA, Departamento de F{\'i}sica, Instituto Superior T{\'e}cnico – IST, Universidade de Lisboa – UL, Avenida Rovisco Pais 1, 1049 Lisboa, Portugal}

\begin{abstract}
In this article, we study Beltrami equilibria for plasmas in near the horizon of a spinning black hole, and develop a framework for constructing the magnetic field profile in the near horizon limit for Clebsch flows in the single-fluid approximation. We find that the horizon profile for the magnetic field is shown to satisfy a system of first-order coupled ODEs dependent on a boundary condition for the magnetic field. For states in which the generalized vorticity vanishes (the generalized `superconducting' plasma state), the horizon profile becomes independent of the boundary condition, and depend only on the thermal properties of the plasma. Our analysis makes use of the full form for the time-independent Ampere's law in the 3+1 formalism, generalizing earlier conclusions for the case of vanishing vorticity, namely the complete magnetic field expulsion near the equator of an axisymmetric black horizon assuming that the thermal properties of the plasma are symmetric about the equatorial plane. For the general case, we find and discuss additional conditions required for the expulsion of magnetic fields at given points on the black hole horizon. We perform a length scale analysis which indicates the emergence of two distinct length scales characterizing the magnetic field variation and strength of the Beltrami term, respectively.
\end{abstract}

\maketitle

\section{\label{sec:level1}Introduction}
Magnetic fields play an important role in the formation and evolution of many astrophysical compact objects such as white dwarfs, neutron stars and black holes \cite{komissarov2007meissner,vietri2008foundations,bhattacharjee2015novel,asenjo2013generating,BicakDvorak1980,Kingetal1975,Wald1974,kunz2017magnetized,gurlebeck2017meissner,gurlebeck2018meissner,pu2017observable}. Black holes are of particular interest because they possess a horizon, distinguishing them from other types of compact objects; the presence of a horizon can in principle lead to qualitatively different behaviors in a surrounding plasma \cite{koljonen2015connection,palenzuela2010magnetospheres,takahashi1990magnetohydrodynamic,nitta1991effects,narayan2008advection,chou1999dynamics,narayan1997advection,parfrey2019first,blandford1977electromagnetic}. One might expect the extreme conditions near the black hole horizon to lead to extreme conditions in the plasma itself so that the large scale magnetic field structure is dominated by plasma dynamics in the vicinity of the horizon. 

Even in the absence of a horizon, a strong coupling between magnetic field and plasma dynamics is already relevant for many situations such as plasma confinement, solar physics, laser-plasma interaction, and magnetic reconnection. An interesting example is the tendency of a magnetized plasma to evolve towards equilibrium states characterized by ordered large scale magnetic field and flow structure. One such state, first derived in the context of single fluid magnetohydrodynamics, $\vec\nabla\times\vec{B}=\Lambda \vec B$ is known as a `relaxed state' where $\vec{B}$ and $\Lambda$ respectively denote the magnetic field and a Lagrange multiplier \cite{woltjer1958theorem,taylor1974relaxation}. Later, these states were extended by incorporating multi-species effects in classical and relativistic plasmas \cite{mahajan1998double,PhysRevLett.90.035001,PhysRevLett.79.3423,steinhauer1998relaxation}. The defining characteristic of these equilibrium states is the alignment of a more general physical quantity known as vorticity $\vec \Omega=\vec B+m/q \, \vec\nabla\times\vec{v}$, with the flow field $v$, in particular, the generalized Beltrami condition $\vec \Omega = \lambda \vec{v}$ (where $\lambda$ is a Lagrange multiplier), which is an equilibrium solution of the time-independent vorticity equation \cite{bhattacharjee2015magnetofluid}. It should be noted here that $\vec\Omega=\vec \nabla \times \vec{P}$, where $\vec P=\vec A+m/q \, \vec v$ is the canonical momentum. The full equilibrium also requires simultaneous satisfaction of the Bernoulli condition signifying homogeneous energy distribution. 

One might expect steady-state flows in the near horizon limit for astrophysical black holes which slowly accrete plasma and it is reasonable to model such plasma flows as force-free states which satisfy the Beltrami condition; the existence and implications of a force-free region near the black hole horizon have been studied extensively in theoretical and numerical frameworks\cite{BlandfordZnajek1977,1969ApJ...157..869G,komissarov2004general,komissarov2007meissner,singh2018magnetic}. Similarly, it is physically reasonable to expect a broader class of Beltrami equilibria which incorporates inertia of plasma constituents in the near horizon limit. In this paper, we adopt the vortical formalism of plasma dynamics to investigate these types of equilibria.

On the other hand, these Beltrami states can also terminate because of changes (via reconnection or other non-ideal effects) in field topology. For example, in solar physics, a class of states known as Double Beltrami (DB) states, the superposition of two force free states, is shown to have the characteristics similar to those of active regions in the solar corona \cite{ohsaki2001magnetofluid,ohsaki2002energy,10.1111/j.1365-2966.2010.16741.x}. The breakdown of these DB states in those active regions can lead to catastrophic eruptive events such as coronal mass ejections (CME). In general, the DB equilibrium has more energy available to cause eruptive events than a force-free linear MHD state and the critical energy of the equilibrium is determined by invariants of the system. If the plasma flow near the horizon of a black hole satisfies the Beltrami condition, then the eventual termination of the Beltrami equilibrium may lead to similar eruptive events with observable signatures---such events and their signatures, while interesting, are beyond the scope of the present work and are left for future investigation. It should be mentioned that we regard such phenomena to be independent of the jet-producing mechanisms explored in the literature such as the well known Blandford-Znajek mechanism \cite{BlandfordZnajek1977}.

To investigate Beltrami equilibria in the near horizon limit of a Kerr black hole, we employ the electrovortical formalism
%covariant description for vortical dynamics, known as the `magnetofluid unification' firs
described in \cite{PhysRevLett.90.035001,mahajan2016relativistic}, which was later extended to curved spacetime in \cite{asenjo2013generating,bhattacharjee2015magnetofluid}. The electrovortical formalism reorganizes the equations of a charged fluid into the form $qU_{\mu}\mathbb{M}^{\mu\nu}=0$. The electrovortical tensor $\mathbb{M}^{\mu\nu}$ is an antisymmetric rank-two tensor constructed from the electromagnetic field strength tensor $F^{\mu\nu}$ and the antisymmetrized derivatives of the `temperature-transformed' fluid flow field $\mathcal{G}U^{\mu}$, where $U^{\mu}$ is the fluid four-velocity. The thermodynamic factor $\mathcal{G}$ is a function of temperature $T$ and entropy $\sigma$. This is similar to the ideal Ohm's law in MHD; one can see the force-free structure when writing the covariant equation of motion in its three-dimensional form
\begin{equation}\label{3Deom}
    \Gamma \vec{\mathcal{E}}_G + \vec{U} \times \vec{\Omega}_G = 0,
\end{equation}
\begin{equation}\label{energy}
\vec{U}\cdot\vec{\mathcal{E}}_G=0
\end{equation}
where $\vec{\mathcal{E}}_G$, $\vec{\Omega}_G$, $\Gamma$ are the generalized electric field, generalized magnetic field (generalized vorticity) and Lorentz factor, respectively\cite{mahajan2016relativistic,bhattacharjee2018surveying,bhattacharjee2018superconducting}. What is remarkable about this formalism is that the complicated dynamics of hot relativistic plasmas in curved spacetime has the familiar force-free MHD state, if expressed in suitably constructed variables. Eq. (\ref{3Deom}) also indicates the plasma is frozen to the generalized magnetic field lines and that the generalized helicity $\mathcal{H}=\langle \vec{P}\cdot\vec{\Omega} \rangle$ is a complete invariant of the model for arbitrary thermodynamics. As a result, this formalism yields a broader class of equilibrium states which is inaccessible to traditional fluid theories\cite{bhattacharjee2015beltrami,pino2010relaxed,bhattacharjee2018surveying}.

Recently, one such equilibrium state, obtained by setting $\Omega_G=0$ in Eq.(\ref{3Deom}), has revealed a `Meissner'-like effect, in which the magnetic field is completely expelled from the equator of a black hole event horizon\cite{bhattacharjee2018superconducting}. This equilibrium state also displays a rich interplay between plasma dynamics and general relativity by satisfying the Bernoulli condition $\mathcal{E}_G=0$ which represents the homogeneity of total energy. Since this result is partly a consequence of the spacetime geometry near the horizon, it is natural to ask whether the expulsion of the magnetic field from the event horizon in a black hole is present in a more general equilibrium state. 

In this article, we develop a general framework for studying Beltrami equilibria in the near horizon limit of a spinning (Kerr) black hole. In particular, we construct a series expansion that we use to obtain expressions for the magnetic field profile on the horizon, valid for arbitrary thermodynamics. We study the general properties of these profiles, identifying conditions for magnetic field expulsion at various points on the horizon. We also generalize the result of \cite{bhattacharjee2018superconducting}; our present analysis includes a previously neglected term in Ampere's law---eliminating an implicit assumption---and we find that that the `Meissner'-like effect at the equator of the horizon for the Beltrami states $\Omega_G=0$ is still present even when this term is included.

\section{Kerr geometry}
Here, we describe the Kerr geometry for an uncharged, rotating black hole. The metric components $g_{\mu \nu}$ can be read off from the line element $ds^2=g_{\mu \nu} dx^\mu dx^\nu$, which in Boyer-Lindquist coordinates $(t,~r,~\theta,~\phi)$ [$r$ being a radial coordinate, and $\theta,~\phi$ being the angular coordinates on a spheroid] takes the form \cite{BoyerLindquist1967,CarrollGR}:
\begin{equation} \label{MEBEBH-KerrBLmain}
\begin{aligned}
ds^2 = & - \left(1-\frac{2 G M r}{\Sigma}\right) dt^2 + \frac{\Sigma}{\Delta} dr^2 + \Sigma \> d\theta^2 \\
& - \frac{4 G M r a \sin^2 \theta}{\Sigma} dt  \> d\phi \\
& + \left(r^2 + a^2 + \frac{2 G M r a^2}{\Sigma}\sin^2 \theta\right) \sin^2 \theta \> d\phi^2 ,
\end{aligned}
\end{equation}

\noindent where $M$ is the mass, $a$ is the spin parameter, and:
\begin{equation} \label{MEBEBH-KerrBLdefs}
\begin{aligned}
\Sigma &:= r^2 + a^2 \cos^2 \theta \\
\Delta &:= r^2 - 2 G M r + a^2.
\end{aligned}
\end{equation}

\noindent The event horizon of the black hole is located at one of the roots of $\Delta$, in particular the root given by the following expression:
\begin{equation} \label{MEBEBH-Horizon}
r_H = G \, M + \sqrt{G^2 \, M^2 - a^2}.
\end{equation}

\noindent As is well-known, the Kerr metric becomes singular at the horizon in Boyer-Lindquist coordinates; in particular, $g_{rr} \propto 1/\Delta$, which diverges in the limit $\Delta \rightarrow 0$.

For our analysis, it will be convenient to work in terms of the metric components rather than their explicit expressions as given in the line element. We begin with the form of a stationary and axisymmetric spacetime\cite{Wald,ChoquetBruhat2015}:
\begin{equation}\label{MEBEBH-StationaryLineElement}
ds^2 = - A^2  dt^2 + 2 \, \beta_\phi \, dt \, d\phi + h^2_1 \, dr^2 + h^2_2 \, d\theta^2 + h^2_3  \, d\phi^2 .
\end{equation}
The quantities $A^2$, $\beta_\phi$, $h_1^2$, $h_2^2$, and $h_3^2$ are functions of $r$ and $\theta$ only, and all correspond to the appropriate components of the Kerr metric (\ref{MEBEBH-KerrBLmain}). It is helpful to relate these quantities to the ADM 3+1 formalism \cite{ADM59,*ADM62,Gourgolhoun3+1,baumgarteshapiro2010,alcubierre2008}, in which spacetime is foliated by a family of three-dimensional spacelike hypersurfaces such that each hypersurface $\Sigma_t$ is defined by a constant value for the time coordinate $t$. The metric is decomposed into the ADM variables  $\alpha$ (the lapse function), $\beta^i$ (the shift vector and $\gamma_{ij}$ (the induced metric), which can be identified from the relations
$g_{tt}=-A^2=-\alpha^2+\gamma_{ij} \beta^i \beta^j$, and $g_{0j} = \gamma_{ij} \beta^j$. The explicit expression for $\alpha$ is:
\begin{equation}\label{MEBEBH-AlphaExpression}
\alpha = \sqrt{\frac{\Delta  \Sigma }{\left(a^2+r^2\right) \left(a^2-\Delta +r^2\right)+\Delta  \Sigma }}.
\end{equation} 

\noindent It is helpful to define the quantity $n^\mu=-\alpha g^{\mu \nu} \nabla_\nu t$, which is the unit normal vector to $\Sigma_t$. One can interpret $n^\mu$ to be the four-velocity for observers whose spatial frames are tangent to surfaces of constant $t$, termed Zero Angular Momentum Observers (ZAMOs). One can decompose the four-velocity with respect to this frame, with $U^0=\Gamma/\alpha$, $\vec{U}=\Gamma \vec{V}$, where $\Gamma=1/\sqrt{1-V^2}$ is the Lorentz factor.

Since we will be considering Beltrami states in the near horizon limit, it is appropriate to establish the scaling of geometric quantities as one approaches the horizon. Note that in the near horizon limit $r \rightarrow r_H$, the Lapse function $\alpha$ vanishes, and the metric component $h_1=\sqrt{\Sigma/\Delta}$ diverges. We find the following behavior for $\alpha$ and $h_{1}$ and their derivatives in the near horizon limit by expanding in powers of $s:=\sqrt{|r-r_H|/r_H}$; WLOG, we set $r_H$ to unity.
%   $s = \sqrt{\delta r}$, where $\delta r:=r-r_H$.
Noting that $\partial_r (\cdot) = (1/2s) \partial_s (\cdot) $, one finds to leading order:
%$\delta r:=r-r_H$:
\begin{equation}\label{MEBEBH-ScalingBehavior}
\begin{aligned}
\alpha &= \alpha_{(1)} s + O(s^3)\\
\partial_r \alpha &= \frac{\alpha_{(1)}}{2 s} + O(s^2)\\
\partial_\theta \alpha &= \alpha^\prime_{(1)} \, s + O(s^3)\\
h_1 &= \frac{h_{1 (-1)}}{s} + O(s)\\
\partial_r h_1 &= - \frac{h_{1 (-1)}}{2 s^3} + O(s^{-1})\\
\partial_\theta h_1 &= \frac{h^\prime_{1 (-1)}}{s} + O(s).
\end{aligned}
\end{equation}

\noindent where all coefficients are functions of $\theta$, and the order of the coefficient appears in the parentheses (so that $h_{1,(I)}$ is the coefficient for the $s^I$ term). One can see that the quantity $\alpha \, h_1 \sim \alpha_{(1)} \, h_{1 (-1)}$ remains finite as one approaches the horizon. Another important quantity to consider is $\beta^\phi:=\beta_\phi/h_3^2$, which scales in the following manner:
\begin{equation}\label{MEBEBH-ScalingBehaviorBeta}
\begin{aligned}
\beta^\phi &= b_{0} + \beta^\phi_{(2)} s^2 + O(s^4).
\end{aligned}
\end{equation}

\noindent where $b_0$ is a constant. Though $\beta^\phi$ does not vanish on the horizon, $\partial_\theta \beta^\phi \sim O(s^2)$ vanishes, and one finds that the combination $h_1^2 \partial_\theta \beta^\phi$ remains finite at the horizon. It turns out that in the near horizon limit, it suffices to consider only these general properties of $\alpha$ $h_1$, and $\beta^\phi$. 

One can also establish the scaling behavior for the motion of matter in the near horizon limit. The Kerr metric admits two Killing vectors $\eta=\partial/\partial t$ and $\psi=\partial/\partial \phi$. with components $\eta^\mu=\delta^\mu_0$ and $\psi^\mu=\delta^\mu_3$. Lowering the indices, one obtains:
\begin{equation}\label{MEBEBH-KillingVectorsLowered}
\begin{aligned}
\eta_\mu &= g_{\mu 0}\\
\psi_\nu &= g_{\mu 3}.
\end{aligned}
\end{equation}

\noindent Killing vectors can be used to define\footnote{These are the energy and angular momentum in the sense that for particles falling from infinity, $E$ and $L$ conserved quantities along geodesics that correspond to respective energy and angular momentum at infinity.} the energy $E$ and angular momentum $L$ for a fluid particle with four-velocity $U^\mu$
\begin{equation}\label{MEBEBH-ConservedQuantities}
\begin{aligned}
E &:= -m \, \eta_\mu U^\mu = m \left(\alpha^2 - \beta_\phi^2 /h_3^2 \right) U^0 - m \, \beta_\phi \, U^3\\
L &:= m \, \psi_\mu U^\mu = m \, \beta_\phi \, U^0 + m \, h_3^2 \, U^3.
\end{aligned}
\end{equation}

\noindent Eliminating $U^3$, one may write:
\begin{equation}\label{MEBEBH-ScalingU0a}
\begin{aligned}
L + \frac{h_3^2 \, E}{\beta_\phi} &= \frac{h_3^2 \, \alpha^2}{\beta_\phi} \,U^0 .
\end{aligned}
\end{equation}

\noindent For finite energy and angular momentum, the left-hand side must be finite at the horizon, and since $\alpha^2 \sim {\alpha}^2_{(1)} \, s^2$, then to leading order, $U^0$ must scale as $U^0 \sim {U}^0_{(2)}/s^2$. From the expression $U^0=\Gamma/\alpha$, it follows that $\Gamma$ scales in the following way:
\begin{equation}\label{MEBEBH-ScalingU0b}
\Gamma \sim \frac{\Gamma_{(-1)}}{s},
\end{equation}

\noindent implying that the fluid velocity reaches the speed of light at the horizon. Note that this scaling behavior depends only on the demand that the energy and angular momentum for the fluid particles are finite. Physically, one expects the Lorentz factor ${\Gamma}$ to diverge, since the four-velocities of ZAMOs become null at the horizon, so infalling matter will appear to be moving at velocities close to the speed of light for near-horizon ZAMOs.

\section{\label{sec:level2}Plasma dynamics in Curved spacetime: Grand Generalized Vorticity}
\subsection{Covariant electrovortical formalism}
The dynamics of a multi-species ideal plasma is summarized in the expression $\nabla_{\mu}T^{\mu\nu}=qU_{\mu}F^{\mu\nu}$, where $T^{\mu\nu}=hU^{\mu}U^{\nu}+pg^{\mu\nu}$ and $h=p+\rho$. Here $p$ and $\rho$ are the respective pressure and plasma density. The dynamical equations may be rewritten in the standard form
\begin{equation}
\label{eomfluid}
mnU^{\nu}\nabla_{\nu}\left(\mathcal{G}U^{\mu}\right)=qnF^{\mu\beta}U_{\beta} - \nabla^{\nu}p ,
\end{equation}
where the quantities $m$, $q$ and $n$ are the respective mass, charge and number density for the constituent particles of the fluid. The fluid four-velocity for each species may be written as $U^{\mu}={dx^{\mu}}/{d\tau}$, where $\tau$ is the proper time for a fluid element. The thermodynamic factor $\mathcal{G}$ is given by the expression $h=mn\mathcal{G}$ with $h$ and $\rho$ being the respective enthalpy and mass density of the fluid. We assume pressure $p=nkT$ and a local Maxwellian distribution function for which the corresponding thermodynamics factor has the form $\mathcal{G}=K_3(mc^2/k_bT)/K_2(mc^2/k_bT)$, where $K_j$ is the modified Bessel function of order $j$ and $k_b$ is the Boltzmann constant.
Though one typically requires a multifluid model to fully describe the behavior of a plasma, we explore a simplified model (which is nonetheless still more general than that of MHD) in which the behavior of a quasineutral plasma is described by the dynamics of an effective single charged fluid in a neutralizing background, assuming that the motion of the effective fluid does not differ strongly from the bulk motion. In a low density and low collisionality plasma where the species have different thermodynamics, it may be important to analyze the dynamics of single species separately in a neutralizing bulk plasma\cite{10.1093/mnras/stv2084}.

The electrovortical formalism is based on the following observations:
\begin{itemize}
    \item An anti-symmetric flow field tensor can be obtained by incorporating temperature into the flow, expressed as 
    $S^{\mu \nu}:= \nabla ^{\mu}\left(\mathcal{G}U^{\nu}\right)
    -\nabla^{\nu}\left(\mathcal{G}U^{\mu}\right)$. 
    Eq.~(\ref{eomfluid}) can then be rewritten as
\begin{equation}
\label{eommagnetofluid}
    qU_{\mu}\mathcal{M}^{\mu\nu}=T\nabla^{\nu}\sigma,
\end{equation}
where $\mathcal{M}^{\mu\nu}=F^{\mu\nu}+\left(m/q\right)S^{\mu\nu}$ is the electrovortical tensor, and the entropy density $\sigma$ for the fluid obeys $T\nabla^{\nu}\sigma = \left(mn\nabla^{\nu}\mathcal{G}-\nabla^{\nu}p\right)/n$\cite{PhysRevLett.90.035001}.

\item A relativistic perfect fluid is isentropic,
\begin{equation}
\label{isentropic}
    U_{\nu}\nabla^{\nu}\sigma=0,
\end{equation}
the entropy density $\sigma$ being constant along a flow line.
\end{itemize}

\noindent Equation (\ref{eommagnetofluid}) for plasma dynamics can be recast in a source free form by defining the following electrovortical potential $\mathbb{P}^\mu$:
\begin{equation}\label{GGVpotential}
    \mathbb{P}^{\mu}=A^{\mu}+\frac{m}{q} \, \mathcal{G} \, U^{\mu} + \sigma \, \nabla^\mu \mathcal{Q} ,
\end{equation}
where $A^{\mu}$ is the usual electrodynamical four-potential and $\mathcal{Q}$ is a scalar, which is defined by the expression
\begin{equation}\label{ZandTemperature}
U_\nu \nabla^\nu \mathcal{Q} = T / q ,
\end{equation}
\noindent and an appropriate set of initial data for $\mathcal{Q}$ which is determined by invariants of the vortical dynamics such as circulation, helicity etc. The covariant equation of motion Eq. (\ref{eommagnetofluid}) may then be written\cite{bhattacharjee2018superconducting}: 
\begin{equation}\label{grandequation}
    qU_{\mu}\mathbb{M}^{\mu\nu}=0,
\end{equation}
where we define the grand electrovortical tensor $\mathbb{M}^{\mu\nu}$ to be
\begin{equation}\label{grandevtensor}
    \mathbb{M}^{\mu\nu}=\nabla^{\mu}\mathbb{P}^{\nu}-\nabla^{\nu}\mathbb{P}^{\mu}.
\end{equation}
It is straightforward to show (keeping in mind Eq. (\ref{ZandTemperature}) for $\mathcal{Q}$) that one can recover Eq. (\ref{eommagnetofluid}) from Eq. (\ref{grandequation}). The source-free formalism summarized in Eqs. (\ref{GGVpotential}-\ref{grandevtensor}) is referred to as the grand generalized vortical formalism in the literature, and we use the modifier `grand' to refer to quantities constructed from the grand electrovortical tensor $\mathbb{M}^{\mu\nu}$.

In general, flow fields $U^\mu$ satisfying Eq.~(\ref{ZandTemperature}) can be written in the form
\begin{equation} \label{SCSC-Clebsch}
T U^\mu = - q \, \nabla^\mu \mathcal{Q} + b^\mu,
\end{equation}
where the vector $b^\mu$ is orthogonal to $U^\mu$\cite{PhysRevD.2.2762}. To simplify the analysis, we require $b^\mu=0$, so that $T U^\mu =  - q \, \nabla^\mu \mathcal{Q}$. One may recognize this Clebsch flow restriction to be the requirement the the flow is hypersurface-orthogonal (i.e. irrotational); the analysis of equilibria for more general flows ($b^\mu \neq 0$) will be explored in a forthcoming article.

\subsection{3+1 decomposition of the electrovortical formalism}
It will be useful to rewrite the covariant formulas of the electrovortical formalism in terms of more familiar three-dimensional variables. We do this by rewriting Eq. (\ref{grandequation}) using the $3+1$ ADM formalism discussed earlier---this decomposition of the electrovortical equations is discussed in detail in \cite{bhattacharjee2015magnetofluid,bhattacharjee2018surveying}. Equation (\ref{grandequation}) is split into space and time components; the spatial components, which form the three-dimensional equation of motion, are obtained by applying the projection operator $\gamma{^\mu}{_\nu}=\delta^{\mu}_{\nu}+n^{\mu}n_{\nu}$; one obtains: 
\begin{equation} \label{SCSC-EoMgen1}
\Gamma (\alpha\vec{\mathcal{E}}_G +\vec\beta\times\vec\Omega_G)+ \vec{u} \times \vec{\Omega}_G = 0,
\end{equation}
where the grand generalized electric field $\vec{\mathcal{E}}_G$ and vorticity $\vec{\Omega}_{G}$ given by the respective equations:
\begin{align}\label{generalefield}
    \vec{\mathcal{E}}_G= \ & \vec{E}-\frac{m}{\alpha q}\vec{\nabla}(\alpha T \mathcal{G}^\prime\Gamma)
    -\frac{m}{q}\left[2\underline{\underline{{\sigma}}}\cdot(\mathcal{G}^\prime T \vec{U})+\frac{2}{3} K \mathcal{G}^\prime T \vec{U}\right] \notag\\
\ &-\frac{m}{q\alpha}\left(\partial_t(\mathcal{G}^\prime T \vec{U}) -\mathcal{L}_{\vec{\beta}}(\mathcal{G}^\prime T  \vec{U})\right);
    \end{align}
\begin{equation} \label{SCSC-GGV}
   \vec \Omega_{G}=\vec{\nabla}\times \vec{\mathbb{P}}_{G}=\vec{\nabla}\times\left(q \vec{A} + m \mathcal{G}^\prime T \vec{U}\right),
\end{equation}
where $\Gamma=1/\sqrt{1-V^2}$ is the Lorentz factor, flow velocities $\vec{U}=\Gamma \vec{V}$ \& $\vec{u}=\Gamma \vec{v}$, $\vec A$ is the vector potential and $\mathcal{G}^\prime=(\mathcal{G}/T-\sigma/m)$ is the modified thermodynamic factor. Here, $\vec{\beta}$ is the shift vector, $\mathcal{L}_{\vec{\beta}}$ is the Lie derivative with respect to $\vec{\beta}$, $\underline{\underline{{\sigma}}}$ is a trace-free rank-2 tensor (with components $ \sigma{^i}{_j}$ formed from $\partial_t{\gamma}_{ij}$ and $\beta^i$) called the shear tensor, and $K$ is the mean curvature for $\Sigma_t$. The dynamics of a hot, relativistic, magnetized plasma as expressed by Eq.~(\ref{SCSC-EoMgen1}) has a similar structure of the ideal Ohm's law in ideal MHD $qU_{\mu}F^{\mu\nu}=0$. It is clear that plasma inertia plays a critical role in the electrovortical dynamics and the characteristics of any equilibria near the Kerr black hole will be fundamentally different. 

\section{The Beltrami Equilibrium}
\subsection{Beltrami equilibria in axisymmetric spacetimes}
Eq.~(\ref{SCSC-EoMgen1}) is trivially satisfied if in the stationary state the combination $\vec{\mathcal{E}_G}+\vec\beta\times\vec\Omega$ equals zero and the grand generalized vorticity $\vec{\Omega}_G$ (GGV) is parallel to the coordinate flow velocity $\vec{u}:=\alpha\vec{U}-\Gamma \vec{\beta}$
\begin{align}\label{Superstate} 
     \alpha\vec{\mathcal{E}}_G +\vec\beta\times\vec\Omega & =\nabla \Psi=0
     \end{align}
     \begin{align}\label{superstate 2}
     \vec{\Omega}_G=q \, \vec{B} + m \, T \, \vec{\nabla} \mathcal{G}^\prime \times \vec{U}=\mu\hat{n}\vec{u},
\end{align}
where the magnetic field is $\vec{B}=\vec{\nabla}\times\vec{A}$, $\mu$ is the separation constant (inverse length) and $\Psi$ contains all the gradient forces.  We have used the vector identity $\vec\nabla\times\vec\nabla \mathcal{Q}=0$ in Eq.(\ref{superstate 2}). The first equation is a general relativistic Bernoulli's condition in Kerr spacetime signifying the homogeneity of total energy, and the second equation is the Beltrami condition. It should be noted here that Eq.(\ref{Superstate}) follows from the steady-state generalized Faraday's law for electrovortical variables with the condition $\nabla\cdot\beta=0$ and $\gamma^{ij}\dot{\gamma}_{ij}=0$~\cite{bhattacharjee2018surveying}. The appearance of $\vec{\beta}$ (implicit in $\vec{u}=\alpha\vec{U}-\Gamma \vec{\beta}$) on the right-hand side of the second equation \eqref{superstate 2} follows from the condition that the plasma is stationary with respect to the Killing vector $\partial/\partial t$.
The separation constant $\mu$ plays the role of a Lagrange multiplier if one were to derive Eq.(\ref{superstate 2}) via a constrained energy minimization principle---the separation constant is related to the invariants of the system such as energy, helicity, etc \cite{mahajan2015multi}. The helicity is useful in understanding astrophysical dynamos, solar wind, fusion as well as determining the conditions for the loss of Beltrami equilibria that lead to eruptive events \cite{ohsaki2001magnetofluid,ohsaki2002energy}. When satisfied, Eqs. (\ref{Superstate}) \& (\ref{superstate 2}) constitute a class of plasma states known as Beltrami-Bernoulli equilibria. The factors in the RHS of Eq. (\ref{superstate 2}) satisfies the continuity equation $\nabla\cdot(\hat{n}\vec u)=0$ where $\hat{n}$ is the density envelope of the plasma species and also ensures that $\vec\nabla\cdot\vec\Omega_G=0$. 

Since we seek solutions corresponding to a steady-state charge neutral Maxwell-fluid system (for example, an electron plasma with ions as the neutralizing background), Eq. (\ref{superstate 2}) should be coupled with the steady state Ampere's law \cite{thorne1982electrodynamics,baumgarteshapiro2010}
\begin{equation}\label{Amph}
    \vec{\nabla}\times(\alpha \vec{B})=4\pi nq\alpha\vec{U} - \pounds_\beta \vec{E}.
\end{equation}

\noindent The term $\pounds_\beta \vec{E}$ in Ampere's law is often implicitly assumed to vanish. However one should not neglect it when considering plasma dynamics in the near horizon limit for generic flows.\footnote{Here, we can establish conditions under which one can neglect the $\pounds_\beta \vec{E}$ term, as was done in \cite{bhattacharjee2018superconducting}. Assuming $\partial_\phi E^i = 0$ (which follows from axisymmetry), one has $\pounds_\beta \vec{E}=-[\vec{E} \cdot \vec{\partial}] \vec{\beta}$. Since the only nonzero component of $\vec{\beta}$ is $\beta^\phi$, $\pounds_\beta \vec{E}$ vanishes if $\vec{E} \cdot \vec{\nabla} \beta^\phi = 0$. Since $\partial_\theta \beta^\phi \rightarrow 0$ in the near horizon limit, axisymmetry demands that $\vec{\nabla} \beta^\phi$ must point in the $r$ direction. Furthermore, for stationary and axisymmetric states, lies in the $r$-$\theta$ plane, so the condition $\pounds_\beta \vec{E}=0$ demands that $\vec{E}$ must point in the $\theta$ direction. Assuming quasineutrality in the comoving frame $\vec{E}=-\vec{V}\times \vec{B}$ and it follows that the flow velocities $\vec{V}$ and magnetic fields $\vec{B}$ must be restricted to the $r$-$\phi$ plane. } The Lie derivative term may be written in terms of partial derivatives as:
\begin{equation}\label{LieDerivativeTerm}
    \pounds_\beta E^i = \beta^j \partial_j E^i - E^j \partial_j \beta^i .
\end{equation}

\noindent Since the shift vector $\vec{\beta}$ is aligned with the $\phi$ direction, the first term vanishes by axisymmetry. However, the second term does not in general vanish. If the plasma is assumed to be quasineutral in the comoving frame of the effective fluid, $\vec{E}=-\vec{U}\times\vec{B}/\Gamma$, so that Ampere's law takes the form:
\begin{equation}\label{Amphmod}
    \vec{\nabla}\times({\alpha \vec{B}_c})=\frac{\hat{n}}{\lambda^2}\alpha \vec{U} - \frac{1}{\Gamma}\left[(\vec{U}\times\vec{B}_c)\cdot\vec{\partial} \right]\vec{\beta},
\end{equation}

\noindent where the fields have been normalized in terms of the cyclotron frequency $B_c=q/m \, B$, $n= \hat{n} n_0$ where $n_0$ is the average density, and the skin depth $\lambda = \sqrt{4 \pi n_0 q^2/m}$, associated with some average density, is an intrinsic length scale of the dynamics. Since Eq. \eqref{Amphmod} is an algebraic equation for $\vec{U}$, one can in principle solve it for the components of $\vec{U}$, and use it in Eq. \eqref{superstate 2} to obtain an equation for the magnetic field $\vec{B}$. One property that simplifies the analysis is the fact that the last term on the RHS of Eq. \eqref{Amphmod} contributes only to the $\phi$ component; the $r$ and $\theta$ components of $\vec{U}$ can then be obtained independently of the $\phi$ component. An explicit calculation reveals that $\vec{U}$ obtained in this way satisfies the steady-state continuity equation $\vec{\nabla}\cdot(\hat{n}\vec{u})=0$.

To simplify the results, it is helpful to define the rescaled components of the magnetic field:
\begin{equation}\label{BBDefs}
\begin{aligned}
    \mathbb{B}_r &:= \alpha  \, h_1 \, B_r\\
    \mathbb{B}_\theta &:= \alpha  \, h_2 \, B_\theta\\
    \mathbb{B}_\phi &:= \alpha  \, h_3 \, B_\phi,
\end{aligned}
\end{equation}

\noindent where $B_r$, $B_\theta$, $B_\phi$ form the orthonormal basis components of $\vec{B}$. Ampere's law yields following components of $\vec{U}$ in the orthonormal basis:
\begin{equation}\label{Ur}
    U_r = q \, \lambda^2 \, \Gamma \, h_1 \, \left[\frac{\alpha \, h_3 \, \partial_\theta \mathbb{B}_\phi}{\Psi} \right]
\end{equation}
\begin{equation}\label{Utheta}
    U_\theta = - q \, \lambda^2 \, \Gamma \, h_2 \, \left[\frac{\alpha \, h_3 \, \partial_r \mathbb{B}_\phi}{\Psi} \right]
\end{equation}
\begin{equation}\label{Uphi}
\begin{aligned}
     U_\phi =& - \frac{ q \lambda ^2 \alpha \Gamma h_3^2 }{\Psi  \left(h_3^3 \lambda ^2 q \left(\mathbb{B}_{\theta } \partial_r \beta^{\phi }-\mathbb{B}_r \partial_\theta \beta^{\phi }\right)+\Psi \right)} \\
     & \times \biggl[
    h_3 \lambda ^2 q \mathbb{B}_{\phi } \left((h_1^2 \partial_\theta \beta^{\phi })  \partial_\theta \mathbb{B}_\phi
    + (h_2^2 \partial_r \beta^{\phi }) \partial_r \mathbb{B}_\phi \right) \\
    & \qquad + \Psi \left(\partial_\theta \mathbb{B}_r - \partial_r \mathbb{B}_\theta \right)
    \biggr],
\end{aligned}
\end{equation}

\noindent where the following quantity has been defined (note that, assuming the number density remains finite, these quantities remain finite in the near horizon limit):
\begin{equation}\label{UphiDefs}
    \Psi := 4 \pi \, m \, \hat{n} \, h_2 \, h_3^2 \, \alpha^2 \, \Gamma \, h_1 .
\end{equation}

\noindent Plugging these results into \eqref{Superstate}, the rescaled magnetic field components become:
\begin{equation}\label{Br}
    \mathbb{B}_r = \alpha \, h_1 \, h_3 \left(\frac{\lambda^2 \, \mu \, \hat{n} \, \Gamma \, \alpha^2 \, h_1 \, \partial_\theta \mathbb{B}_\phi}{\Psi } - \frac{m \, T \, \mathcal{G}^\prime_{,\theta}}{q \,h_2} U_\phi\right),
\end{equation}
\begin{equation}\label{Btheta}
    \mathbb{B}_\theta = \alpha \, h_2 \, h_3 \left(\frac{\lambda ^2 \, \mu \, \hat{n} \, \Gamma \, \alpha ^2 \, h_2 \, \partial_r \mathbb{B}_\phi}{\Psi }
    - \frac{m \, T \, \mathcal{G}^\prime_{,r}}{q \, h_1} U_\phi\right),
\end{equation}
\begin{equation}\label{Bphi}
\begin{aligned}
    \mathbb{B}_\phi =& \lambda ^2 m \frac{\alpha^2 \, \Gamma \, h_3^2 \, T \left[h_1^2 \, \mathcal{G}^\prime_{,\theta} \, \partial_\theta \mathbb{B}_\phi + h_2^2 \, \mathcal{G}^\prime_{,r} \, \partial_r \mathbb{B}_\phi \right]}{h_1 h_2 \Psi } \\
    & + \frac{ \mu \, \hat{n} \, \alpha  \, h_3^2 \left(\alpha \, U_\phi - \Gamma \, \beta^\phi \right)}{q},
\end{aligned}
\end{equation}

\noindent where $U_\phi$ is given in \eqref{Uphi} and we have assumed the thermodynamic potentials (in particular $\mathcal{G}^\prime$) depend only on $r$ and $\theta$ and have defined $\mathcal{G}^\prime_{,\theta}=\partial \mathcal{G}^\prime/\partial \theta$ and $\mathcal{G}^\prime_{,r}=\partial \mathcal{G}^\prime/\partial r$.

Eqs. (\ref{Br}-\ref{Bphi}) form the complete description of Beltrami states of an ideal plasma in a spacetime given by the line element (\ref{MEBEBH-StationaryLineElement}). Next, we analyze the characteristics of these equations in the near horizon limit of a black hole.

\subsection{Near horizon limit}
We now consider the behavior of magnetic fields in the near horizon limit. As discussed in \cite{thorne1982electrodynamics}, the tangential components of the magnetic field ${B}_\theta$ and ${B}_\phi$ (in the orthonormal frame) diverge in the near horizon limit---this is attributed to the fact that the four-velocities of ZAMOs become null at the horizon. On the other hand, the rescaled components $\mathbb{B}_r$,  $\mathbb{B}_{\theta}$ and $\mathbb{B}_{\phi}$ remain finite or are zero in the near horizon limit; for this reason, it is appropriate to work in terms of these components (which are lowered-index coordinate basis elements).

To perform the near horizon analysis, we expand in $s=\sqrt{|r-r_H|/r_H}$. For some function $Q(r,\theta)$, the expansion will be denoted in the following manner:
\begin{equation}\label{QExpansionNotation}
Q = \sum_I \, Q_{(I)} \, s^I
\end{equation}

\noindent where $Q_{(I)}$ are functions of $\theta$ only. In general, leading order terms can have inverse powers of $s$ ($I$ can have negative integer values), and linear terms in $s$ can yield divergent radial derivatives at the horizon by virtue of $\partial_r (\cdot) = (1/2s) \partial_s (\cdot) $. To simplify the analysis, the thermodynamic potentials and their derivatives are assumed to be finite in the near horizon limit, meaning that we exclude inverse and odd powers of $s$.

Equations (\ref{Ur}-\ref{Uphi}) can be used to place limits on the leading order and odd powers for the magnetic field. First, recall that $\vec{U}= \Gamma \vec{V}$, and that (as shown earlier) ${V}^2=1$ at the horizon for finite energy and angular momentum. Then, making use of 
\eqref{MEBEBH-ScalingBehavior} and \eqref {MEBEBH-ScalingBehaviorBeta}, and the fact that $\alpha \, h_1$ $h_1^2 \partial_\theta \beta^\phi$ are finite at the horizon, we find from Eqs. (\ref{Ur}-\ref{Uphi}) that the radial derivatives $\partial_r \mathbb{B}_{\theta}$, $\partial_r \mathbb{B}_{\phi}$ and the derivative $\partial_\theta \mathbb{B}_r$ can diverge no faster than $s^{-1}$, which implies that the smallest possible term of odd power is $O(s)$.

We are now in a position to examine the leading order behavior for Eq. \eqref{Btheta} for $\mathbb{B}_\theta$. From the considerations discussed in the preceding paragraph, one can show that the terms within the parentheses of Eq. \eqref{Btheta} cannot have inverse powers of $s$ to leading order. The overall factor of $\alpha$ implies $\mathbb{B}_\theta \rightarrow 0$ on the horizon, or that the coefficients in the expansion for $\mathbb{B}_\theta$ satisfy $\mathbb{B}_{\theta (I)}=0$ for $I \leq 0$.

Now we consider the expansion of Eqs. (\ref{Br}-\ref{Bphi}) in $s$; in principle, one can solve these equations in the near horizon limit by demanding that the equations hold to each order in $s$. The full expansion, performed in \textit{Mathematica}, is rather complicated, and will not be presented here in full. However, we will describe some relevant features of the expansion. Upon Eq. \eqref{Br} for $\mathbb{B}_r$, we find that the leading order term is $O(s^{-1})$, which diverges at the horizon. This divergent term can be removed with the condition:
\begin{equation}\label{angB1cond}
\mathbb{B}_{\theta (1)} = 
\frac{q \lambda^2 h_{2 (0)} \beta^\phi_{(2)} \mathbb{B}_{\phi (0)} \mathbb{B}_{\phi (1)}}{4 m \pi \hat{n}_{(0)} h_{3 (0)} \alpha_{(1)}^2 h_{1(-1)} \Gamma_{(-1)}}.
\end{equation}

\noindent which can be satisfied if $\mathbb{B}_{\theta (1)}=\mathbb{B}_{\phi (1)}=0$. When expanding Eq. \eqref{Btheta} for $\mathbb{B}_\theta$, we find that the first order ($O(s)$) term implies $\mathbb{B}_{\theta (1)}=0$ (as argued earlier, the zeroth order ($O(s^0)$) term implies $\mathbb{B}_{\theta (0)}=0$). The left hand side of Eq. \eqref{angB1cond} vanishes, which implies that either $\mathbb{B}_{\phi (0)}=0$ or $\mathbb{B}_{\phi (1)}=0$. If the fluid is flowing into the horizon (as one might expect), then Eq. \eqref{Ur} indicates that a nonzero component $V_r$ requires $\partial_\theta \mathbb{B}_{\phi} \neq 0$, which in turn implies $\mathbb{B}_{\phi (0)} \neq 0$. It follows that $\mathbb{B}_{\phi (1)}=0$, meaning that $\partial_r\mathbb{B}_{\phi}$ must be finite at the horizon. This, combined with Eq. \eqref{Utheta} implies that the fluid velocity $\vec{V}$ must lie in the $r$-$\phi$ plane; since $\vec{B}$ also lies in the $r$-$\phi$ plane, one can show that $\pounds_{\vec{\beta}}\vec{E} \rightarrow 0$ at the horizon as was assumed in \cite{bhattacharjee2018superconducting}.

The expression for $\mathbb{B}_{\phi (0)}$ may be obtained from the $O(s^0)$ term in Eq. \eqref{Bphi} for $\mathbb{B}_\phi$, which yields a differential equation for $\mathbb{B}_{\phi (0)}(\theta)=X(\theta)$ of the form;
\begin{equation}\label{angODE}
\frac{\partial X(\theta)}{\partial \theta} = 
\Lambda(\theta) + \Phi(\theta) X(\theta),
\end{equation}

\noindent which admits the solution 
\begin{equation}\label{HorizonBsoln}
\begin{aligned}
X(\theta) = &
 \exp\left\{\int_{\theta_0}^{\theta}\Lambda(x) dx\right\} \\
 &\times\left[C_1+\left[\int_{\theta_0}^{\theta} \frac{\Phi(y)} {\exp\left\{\int_{\theta_0}^{z}\Lambda(z) dz\right\}}dy\right]\right] ,
 \end{aligned}
\end{equation}

\noindent where:
\begin{equation}\label{Fsphi}
\begin{aligned}
\Lambda(\theta) & = \frac{4 \pi \mu \beta^\phi_{(0)} \alpha _{(1) } \Gamma _{(-1) } h_{2 (0) }^2 h_{3 (0) }^2 \hat{n}_{(0) }^2}{\lambda ^2 q T_{(0) } \partial_\theta \mathcal{G}^\prime_{(0)}}\\
\Phi(\theta) & = \frac{4 \pi  h_{2 (0) }^2 \hat{n}_{(0) } }{\lambda ^2 T_{(0) } \partial_\theta \mathcal{G}^\prime_{(0)}}.
\end{aligned}
\end{equation}

\noindent Expanding \eqref{Bphi} to $O(s)$, making use of the results $\mathbb{B}_{\theta (1)}=\mathbb{B}_{\phi (1)}=0$, yields the constraint $ \mu \hat{n}_{(0)} \Gamma_{(0)}=0$, which can be satisfied if $\mu=0$ or $\Gamma_{(0)}=0$; for general $\mu$, we set $\Gamma_{(0)}=0$; we remind the reader that the leading order behavior in $\Gamma$ comes from $\Gamma_{(-1)}$, not $\Gamma_{(0)}$.

A first order differential equation for $\mathbb{B}_{r (0)}$ can be obtained from the $O(s^0)$ term in Eq. \eqref{Br}, which is similar in form to Eq. \eqref{HorizonBsoln}, but now depends on $\mathbb{B}_{\theta (2)}$ and $\mathbb{B}_{\phi (2)}$. In order to obtain expressions for $\mathbb{B}_{\theta (2)}$ and $\mathbb{B}_{\phi (2)}$, Eqs. \eqref{Btheta}, and \eqref{Bphi} must be expanded to $O(s^2)$ and the equation $\vec{\nabla}\cdot\vec{B}=0$ must be expanded to $O(s)$. The $O(s^2)$ term in Eq. \eqref{Bphi} yields an algebraic constraint:
\begin{equation}\label{Bconstraint}
\begin{aligned}
q \mathbb{B}_{\theta  (2) } &=
\frac{\lambda ^2 \mu  q \alpha _{(1) } \mathcal{G}^\prime_{(2) } h_{2 (0) } \partial_\theta \mathbb{B}_{\phi  (0) }}{4 \pi  m h_{(-1) } h_{3 (0) } \partial_\theta \mathcal{G}^\prime_{(0)}}
-\frac{q \mathcal{G}^\prime_{(2) } h_{2 (0) }^2 \mathbb{B}_{r (0) }}{h_{(-1) }^2 \partial_\theta \mathcal{G}^\prime_{(0)}}\\
&-\frac{\lambda ^2 \mu  q \alpha _{(1) } h_{2 (0) } \mathbb{B}_{\phi  (2) }}{4 \pi  m h_{(-1) } h_{3 (0) }}
\end{aligned}
\end{equation}

\noindent The $O(s^0)$ term in Eq. \eqref{Br}, combined with the $O(s)$ terms in $\vec{\nabla}\cdot\vec{B}=0$ (the $O(s^0)$ term yields $\mathbb{B}_{r (1)}=0$) and $O(s^2)$ terms in Eq. \eqref{Bphi} yields a system of linear first order coupled ODEs of the form:
\begin{equation}\label{Bsystem}
\partial_\theta \vec{Z} = \vec{\xi} + \Xi \cdot \vec{Z}
\end{equation}

\noindent where $\Xi=\Xi(\theta)$ is a $3\times3$ matrix and the components of $\vec{Z}$ are defined as:
\begin{equation}\label{Zdefs}
\begin{aligned}
Z_1 & := \mathbb{B}_{r (0)} \\
Z_2 & := \mathbb{B}_{\theta (2)} \\
Z_3 & := \mathbb{B}_{\phi (2)} .
\end{aligned}
\end{equation}

\noindent It is known that a system of the form \eqref{Bsystem} yields unique solutions for the initial value problem. However, the equation is not homogeneous, as the components of the vector $\vec{\xi}=\vec{\xi}(\theta)$ do not vanish. The explicit components have the following form:
\begin{equation}\label{xivectdefs}
\begin{aligned}
\xi _1 &= -\frac{\mu  \alpha _{(1) } h_{(-1) } h_{2 (0) } \hat{n}_{(0) } \partial_\theta \mathbb{B}_{\phi  (0) }}{m h_{3 (0) } T_{(0) } \partial_\theta \mathcal{G}^\prime_{(0)}} \\
&\qquad-\frac{\lambda ^2 q \partial_\theta \beta^\phi_{(2)} h_{(-1) } \mathbb{B}_{\phi  (0) } \partial_\theta \mathbb{B}_{\phi  (0) }}{4 \pi  m \alpha _{(1) }^2 \Gamma _{(-1) } h_{2 (0) } h_{3 (0) } \hat{n}_{(0) }} \\
\xi _2 &= -\frac{h_{2 (0) }^2 \mathbb{B}_{r (2) }}{h_{(-1) }^2} \\
\xi _3 &= \partial_\theta \mathbb{B}_{\phi  (0) } \left(-\frac{\partial_\theta \mathcal{G}^\prime_{(2)}}{\partial_\theta \mathcal{G}^\prime_{(0)}}+\frac{\hat{n}_{(2) }}{\hat{n}_{(0) }}-\frac{T_{(2) }}{T_{(0) }}\right)+\frac{2 h_{2 (2)} \partial_\theta \mathbb{B}_{\phi (0)}}{h_{2 (0) }}\\
&\qquad + \frac{4 \pi \mu \hat{n}_{(0) }^2 h_{2 (0) }^2  h_{3 (0) }^2 \alpha _{(1)}  \beta^\phi_{(0)} \Gamma _{(-1) }}{\lambda ^2 q T_{(0)} \partial_\theta \mathcal{G}^\prime_{(0)}}\biggl[\frac{\beta^\phi_{(2)}}{\beta^\phi_{(0)}} + \frac{ \hat{n}_{(2) }}{\hat{n}_{(0) }}\\
&\qquad + \frac{2 h_{3 (2) } }{ h_{3 (0) }} + \frac{\Gamma_{(1)} }{\Gamma _{(-1)}} + \frac{\alpha_{(3)} }{\alpha _{(1)}} \biggr].
\end{aligned}
\end{equation}

\noindent Note that $\xi _2$ depends on $\mathbb{B}_{r (2) }$, which may be interpreted as the value of $\partial_r\mathbb{B}_{r}$ evaluated on the horizon, since $\mathbb{B}_{r (1)}=0$ (which is obtained from the zeroth-order term in $\vec{\nabla}\cdot\vec{B}=0$). To specify $\mathbb{B}_{r (2)}$, one must continue the expansion to higher powers in $s$, which will depend on higher order coefficients; in this sense, $\partial_r\mathbb{B}_{r}$ captures the dependence of the horizon profile on the behavior of the magnetic field far from the horizon, at least for $\mu \neq 0$ (the $\mu=0$ case will be discussed later). It is therefore appropriate to regard $\mathbb{B}_{r (2)}$ as a boundary condition for the magnetic field at the horizon.

For completeness, the nontrivial components of the matrix $\Xi=\Xi(\theta)$ are written below:
\begin{equation}\label{Ximatrixdefs}
\begin{aligned}
\Xi _{11} &= \frac{4 \pi  h_{2 (0) }^2 \hat{n}_{(0) }}{\lambda ^2 T_{(0) } \partial_\theta \mathcal{G}^\prime_{(0)}}
\\
\Xi _{13} &= \frac{\lambda ^2 q \beta^\phi_{(2)} h_{2 (0) } \mathbb{B}_{\phi  (0) }}{4 \pi  m \alpha _{(1) }^2 \Gamma _{(-1) } h_{(-1) } h_{3 (0) } \hat{n}_{(0) }}\\
\Xi _{21} &= \frac{h_{1 (1)} h_{2 (0) }^2 }{h_{(-1) }^3}+\frac{\alpha_{(3)} h_{2 (0) }^2 }{\alpha _{(1) } h_{(-1) }^2}-\frac{h_{3 (2) } h_{2 (0) }^2}{h_{(-1) }^2 h_{3 (0) }} \\
&\qquad - \frac{h_{2 (2) } h_{2 (0) } }{h_{(-1) }^2}\\
\Xi _{22} &= \frac{\partial_\theta \alpha_{(1)} }{\alpha _{(1) }}+\frac{\partial_\theta h_{2 (0) }}{h_{2 (0) }}-\frac{ \partial_\theta h_{3 (0) }}{h_{3 (0) }}\\
&\qquad - \frac{\partial_\theta h_{1 (-1) }}{h_{(-1) }}\\
\Xi _{32} &= -\frac{4 \pi  \mu  \alpha _{(1) } h_{(-1) } h_{2 (0) } h_{3 (0) } \hat{n}_{(0) }^2}{\lambda ^2 m \mathcal{G}^\prime_{(2) } T_{(0) }^2 \partial_\theta \mathcal{G}^\prime_{(0)}}\\
\Xi _{33} &= \frac{\mathcal{G}^\prime_{(2) } h_{2 (0) }^2}{h_{(-1) }^2 \partial_\theta \mathcal{G}^\prime_{(0)}}-\frac{\mu ^2 \alpha _{(1) }^2 h_{2 (0) }^2 \hat{n}_{(0) }^2}{m^2 \mathcal{G}^\prime_{(2) } T_{(0) }^2 \partial_\theta \mathcal{G}^\prime_{(0)}}+\frac{4 \pi  h_{2 (0) }^2 \hat{n}_{(0) }}{\lambda ^2 T_{(0) } \partial_\theta \mathcal{G}^\prime_{(0)}}
,
\end{aligned}
\end{equation}

\noindent with the remaining components being $\Xi _{12}=1$, $\Xi _{23}=0$, $\Xi _{31}=1$. 

One can specify the horizon profile in the $\mu=0$ case without specifying $\mathbb{B}_{r (2) }$. Note that $\Xi _{32}=0$ when $\mu=0$, so that the equation for $\mathbb{B}_{\phi (2) }$ decouples from $\mathbb{B}_{\theta (2)}$, and has the form given in Eq. \eqref{angODE}, with $X=\mathbb{B}_{\phi (2) }$ and the coefficients $\Lambda=\xi_3$,  $\Phi=\Xi_{33}$ evaluated at $\mu=0$. One may then solve the ODE for $\mathbb{B}_{\theta (2)}$ first, then use the constraint \eqref{Bconstraint} to eliminate the dependence on $\mathbb{B}_{\theta (2)}$ in the ODE for $\mathbb{B}_{r (0)}$. The resulting equation also has the form of Eq. \eqref{angODE}, with $X=\mathbb{B}_{r (0) }$ and the coefficients $\Lambda=\xi_1$ (evaluated at $\mu=0$) and:
\begin{equation}\label{CfBr0}
\begin{aligned}
\Phi(\theta) & =\Xi_{11} + \Xi_{13} -\frac{\mathcal{G}^\prime_{(2) } h_{2 (0) }^2 \mathbb{B}_{r (0) }}{h_{(-1) }^2 \partial_\theta \mathcal{G}^\prime_{(0)}}.
\end{aligned}
\end{equation}

\noindent The horizon profile in the $\mu=0$ case can be obtained in a manner independent of the behavior of the magnetic field far from the horizon. It should be noted that the profile depends on the horizon profile of the fluid potentials $\mathcal{G}^\prime$, $\hat{n}$, and also their normal derivatives $\mathcal{G}^\prime_{(2)}$, $\hat{n}_{(2)}$ (note also the dependence on $\Gamma_{(2)}$), can be determined by the Bernoulli condition along with an appropriate equation of state. 

One can still extract some general properties of the magnetic field without specifying the equation of state. As discussed in \cite{bhattacharjee2018superconducting}, one can have $\mathcal{G}^\prime_{,\theta}=0$ at various points on the horizon. In particular, this condition may be satisfied at the equator of the horizon under the assumption that the thermal properties of the fluid are symmetric about the equatorial plane, or at any value of $\theta$ where $\mathcal{G}^\prime$ has a local maximum or minimum . We also add that regularity conditions on the axis of the black hole require $\mathcal{G}^\prime_{,\theta}=0$ must hold at the poles, though one must also account for the fact that $h_{3 (0) }=0$ at the poles as well. At points on the horizon where $\mathcal{G}^\prime_{,\theta}=0$, Eqs. \eqref{Br} and \eqref{Bphi} yield the following expressions:
\begin{equation}\label{Gmax}
\begin{aligned}
\mathbb{B}_{r (0) } &= \mu  \frac{\lambda ^2 \alpha _{(1) } h_{(-1) } \partial_\theta \mathbb{B}_{\phi  (0) }}{4 \pi  m h_{2 (0) } h_{3 (0) }} \\
\mathbb{B}_{\phi  (0) } &= -(\mu/q) b_0 \alpha _{(1) } \Gamma _{(-1) } h_{3 (0) }^2 \hat{n}_{(0) }
\end{aligned}
\end{equation}

\noindent If $\mu=0$, the magnetic field is completely expelled at points on the horizon where $\mathcal{G}^\prime_{,\theta}=0$, recovering the result in \cite{bhattacharjee2018superconducting} and generalizing it to the case where $\pounds_{\vec{\beta}} \vec{E} \neq 0$ (though as in that case, we still consider Clebsch flow). 

In the more general $\mu \neq 0$ case, we argue that the extrema of $\mathbb{B}_{\phi  (0)}$ cannot all coincide with $\mathcal{G}^\prime$. At a extrema, $\partial_\theta \mathbb{B}_{\phi  (0) }=0$, which combined with the expression for $U_r$ in Eq. \eqref{Ur} implies that the fluid velocity is tangent to the horizon. One might expect such behavior for plasma elements to be unphysical near the horizon, since there are no circular orbits close to the horizon for finite $1-a$ ($a$ being the spin parameter) and there is no physical mechanism that can prevent the plasma elements from falling in; one, therefore, expects the density $\hat{n}_{(0) }=0$ to vanish at these points. However, this implies that $\mathbb{B}_{\phi  (0)}$ must vanish as well; if the extrema for $\mathbb{B}_{\phi  (0)}$ all coincide with an extrema of $\mathcal{G}^\prime$, then Eq. \eqref{Gmax} holds for all extrema of $\mathbb{B}_{\phi  (0)}$, and follows from physical considerations that $\mathbb{B}_{\phi  (0)}=0$ everywhere on the horizon. However, this would imply that $U_r=0$, so no plasma falls into the black hole. This indicates that there must exist at least one extrema for $\mathbb{B}_{\phi  (0)}$ that does not coincide with an extremum of $\mathcal{G}^\prime$. This does not, of course, exclude the possibility that \textit{some} extrema for $\mathbb{B}_{\phi  (0)}$ can coincide with an extremum of $\mathcal{G}^\prime$, and in the instances where this occurs, one has a complete expulsion of the magnetic field at these coincident points on the horizon.

Though the Beltrami equilibria we have described here permit nonvanishing magnetic fields (in which energy can be stored), a stability analysis is needed to determine whether these Beltrami equilibria can be used to describe eruptive events similar to those observed in solar plasmas. In particular, whether such equilibria are stable or metastable depends on the exact relationship between helicity and total energy, which has not been discussed in the present analysis. Such an analysis will be left for future work.

\subsection{Length scale analysis}
The main distinction between the electrovortical formalism and that of GRMHD is that the electrovortical formalism contains multiple length scales which allow one to describe in greater detail the dependence of the macroscopic dynamics on the microphysical properties of the plasma. To illustrate this, we extend the length scale analysis presented in \cite{mahajan2016relativistic} to the GR Beltrami states. The generalized superconducting state $\vec\Omega_G=0$ corresponds to the limit $\mu=0$, or the vanishing of the length scale $L_B=\mu\lambda^2$, which in turn corresponds to the scale at which the Beltrami term in the equations becomes important. By defining $\hat{\lambda}^2=\zeta\mathcal{G}^\prime$, we may write Eqs.(\ref{Btheta} \& \ref{Bphi}), as
\begin{align}
    \frac{1}{\hat{\lambda}^2}=\left(\frac{1}{h_1}\frac{\partial \ln{\mathcal{G}^\prime} }{\partial r}\right)\left(\frac{1}{h_1}\frac{\partial \ln{\mathbb{B}_{(\theta,\phi)}} }{\partial r }\right)=\frac{1}{L_g L_{mag}}
\end{align}
where $1/L_g=1/h_1(\partial\ln{\mathcal{G}^\prime}/\partial r)$ and $1/L_{mag}=1/h_{1}(\partial\ln{\mathbb{B}_{(\theta,\phi)}}/\partial r)$. To keep our calculation simple, we are considering the limit when $\theta=\pi/2$ and the thermodynamic factor is symmetric about the equitorial plane.
Therefore, the magnetic field variation occurs on a hybrid length scale $L_{mag}=\hat{\lambda}^2/L_g$ which is a combination of a modified skin depth and a thermal gradient scale. For $L_g > \hat{\lambda}$, the magnetic field can vary at a length scale shorter than skin depth which one might expect for relativistically hot plasmas. In general, the skin depth characterizes the kinetic-magnetic reservoir of energy arising from microscale physics which can drive the dynamo and reverse-dynamo mechanism \cite{mahajan2016relativistic,mahajan2005acceleration,lingam2015modelling}. 

On the other hand, when $L_B\neq{0}$ (in our case, Eq.(\ref{superstate 2})), there are two intrinsic length scales $L_g$ and $L_B$ (associated with the magnetic field structures) in this formalism which do not appear in GRMHD. We note here that the generalized helicity in this formalism is a topological invariant if the divergence of helicity four vector $\mathcal{K}^{\mu}$ is zero i.e. $\nabla_{\mu}\mathcal{K}^{\mu}=0$. Here, we define the four helicity as $\mathcal{K}^{\mu}=\mathbb{M}^{*\mu\nu}\mathbb{P}_{\nu}$ where $\mathbb{M}^{*\mu\nu}$ is the dual of electro-vortic field tensor $\mathbb{M}^{\mu\nu}$. One can compute the divergence as
\begin{equation}
    \nabla_{\mu}\mathcal{K}^{\mu}=\frac{1}{2}\mathbb{M}_{\mu\nu}\mathbb{M}^{*\mu\nu}=-2\vec\Omega_G\cdot\vec{\mathcal{E}}_G=0
\end{equation}
where we have used the Eq.(\ref{SCSC-EoMgen1}) to compute the last equality. 

The corresponding species helicity is written as
\begin{equation} \label{spechelicity}
h=-\int_{\Sigma_t} n_{\mu} \mathcal{K}^\mu \sqrt{\gamma} \, d^3x=\int_{\Sigma_t} \vec{\mathbb{P}}_G\cdot\vec\Omega_G \sqrt{\gamma} \, d^3x
\end{equation}
where $\vec{\mathbb{P}}=\vec{A}+m/q \ \mathcal{G}\vec{U}+\sigma\vec{\nabla} Q$ is the generalized three-momentum, $n^\mu$ is the 4-velocity of the Eulerian observers (the unit normal vector to constant $t$ surfaces $\Sigma_t$) and $\gamma=\det (\gamma_{ij})$.  The Beltrami condition, when substituted from Eq. (\ref{superstate 2}), into Eq.(\ref{spechelicity}) determines the Beltrami length scale $L_B$ completely . 

One can therefore relate $L_B$ to the profile-modified skin depth $\hat{\lambda}^2$, the thermal gradient $L_g$ and helicity $\mathcal{H}$ near the black hole event horizon. One might recognize the similarity to the Taylor relaxed state in MHD where the length scale is fully determined by the ratio between magnetic helicity and magnetic energy. It should be noted that generalized helicity is identically zero since the Beltrami length scale vanishes.
Finally, we remark that the generalized helicity is useful for establishing the criteria for the destabilization of Beltrami equilibria in solar physics \cite{ohsaki2001magnetofluid,ohsaki2002energy}, and may be similarly useful for investigating the stability of Beltrami equilibria in black hole spacetimes.

\section{Summary and Discussion}
In this work, we have described a single fluid Beltrami state for an ideal plasma with Clebsch flow surrounding s rotating black hole. In particular, we have presented a framework for characterizing the behavior of magnetic fields near black hole horizons, valid in Boyer-Lindquist coordinates in which the metric components become singular at the horizon. The Beltrami condition in rotating black holes dictates the alignment of generalized vorticity and coordinate flow velocity which is completely different than non-rotating black holes. The inherent rotation of spacetime also fundamentally alters the generalized Bernoulli's condition which indicates the balances among different potential forces. 

We have demonstrated how one can obtain the magnetic field profile at the horizon, given the profiles for the fluid quantities $\mathcal{G}^\prime$, $T$ and $\hat{n}$. In particular, we find that the tangential profile for the magnetic field at the horizon can be obtained by expanding Eqs. (\ref{Br}-\ref{Bphi}) to zeroth order in $s=\sqrt{|r-r_H|/r_H}$, with $\mathbb{B}_\theta=0$, and $\mathbb{B}_\phi$ given by Eqs. \eqref{HorizonBsoln} and \eqref{Fsphi}. The horizon profile for the radial component $\mathbb{B}_r$ requires expanding Eqs. (\ref{Br}-\ref{Bphi}) to second order in $s$ and $\vec{\nabla} \cdot \vec{B}=0$ to zeroth order; in doing so we find that given the radial derivative $\partial_r\mathbb{B}_r$ at the horizon, $\mathbb{B}_r$ can be obtained from the system of equations described in Eqs. \eqref{Bconstraint}, \eqref{Bsystem}, and \eqref{Zdefs}. We also find that the radial derivatives of the magnetic field at the horizon must be finite. In the $\mu=0$ case, the horizon profile can be obtained without specifying $\partial_r\mathbb{B}_r$ at the horizon, so that the system Eqs. \eqref{Bconstraint}, \eqref{Bsystem}, and \eqref{Zdefs} for $\mathbb{B}_r$ depends only on the profiles for the fluid quantities and the other components of the magnetic field at the horizon. 

We have also described some general features of the horizon profile, and have extended the analysis of \cite{bhattacharjee2018superconducting} for the generalized `superconducting' states (in the sense of vanishing generalized vorticity) to a more general class of flows in which $\pounds_{\vec{\beta}}\vec{E}$ is not assumed to be zero a priori (though this holds at the horizon), finding that the expulsion of magnetic fields at the maxima of the thermal factor $\mathcal{G}^\prime$ on the horizon holds for these more general flow conditions. We have also demonstrated for more general Beltrami states satisfying $\mu \neq 0$, one can also have magnetic field expulsion at points where the extrema of the horizon profiles for $\mathcal{G}^\prime$ and $\mathbb{B}_\phi$ coincide and have argued that $\mathbb{B}_\phi$ must possess extrema that do not coincide with those of $\mathcal{G}^\prime$. 

Of course, a complete account of the magnetic field in the near horizon limit requires an expansion to higher order in $s=\sqrt{|r-r_H|/r_H}$, at least to the point needed to determine $\mathbb{B}_r$. Furthermore, knowledge of the thermal profiles for the fluid, in particular the form of $\mathcal{G}^\prime$, $\hat{n}$, $T$ and their radial derivatives at the horizon must be supplied to obtain an explicit profile for the magnetic field; in principle, this can be obtained by specifying an appropriate equation of state. This is because the Beltrami states are essentially solutions to the fluid equations---the magnetic field profiles we obtained in this way are those which correspond to a Clebsch flow Beltrami state for all thermodynamics. The most remarkable aspect of this formalism is the emergence of two length scales i.e. thermodynamics modified intrinsic length $L_{mag}$ and Beltrami length $L_B$ which is foreign to the existing single or multi-fluid plasma models. These length scales are also related and completely determined by the generalized helicity which is a conserved quantity of the system. 
In this sense, the results we have obtained here are rather general---the main assumptions of our model are Clebsch flow and the single fluid approximation.

\section{Acknowledgments}
We thank Filip Hejda and David J. Stark for their feedback on this work. We also thank Vitor Cardoso and David Hilditch for useful comments. JCF acknowledges support from FCT grant number PTDC/MAT-APL/30043/2017. Funda{\c{c}}{\~a}o para a Cie{\^n}cia e Tecnologia Award/Contract Number UIDB/00099/2020.

\section{Data availability}
The data that support the findings of this study are available from the corresponding author upon reasonable request.

\bibliographystyle{unsrt}
\bibliography{ref_new} 

\end{document}